\documentclass[twocolumn,showpacs,preprintnumbers,amsmath,amssymb]{revtex4}
\usepackage{amssymb}
\usepackage{txfonts}
\usepackage{bbm}
\usepackage{amsfonts}
\usepackage{mathrsfs}
\usepackage{amsmath}
\usepackage{graphicx}

\begin{document}

\title{Effect of spin excitations on the property of quasiparticles in electron-doepd cuprates}

\author
{Tao Zhou and C. S. Ting} \affiliation{Texas Center for
Superconductivity and Department of Physics, University of Houston,
Houston, Texas 77204}

\date{\today}

\begin{abstract}

It is proposed that the $50-70$ meV dispersion anomaly (kink) in
electron-doped cuprates revealed by recent angle-resolved
photoemission spectroscopy experiments is caused by coupling with
the spin fluctuation. We elaborate that the kink exists both along
nodal and antinodal directions, and both in the superconducting and
normal state. The renormalized effect for the density of states is
also studied and the hump feature outside the superconducting
coherent peak is established, consistent with recent scanning
tunnelling microscopy experiments.

\end{abstract}

\maketitle

Although high-temperature superconductivity in cuprates was
discovered more than twenty years ago, the mechanism of their
unusually high critical temperatures has not yet been
clarified~\cite{leea}. An insightful view may be obtained through
the understanding of the role played by certain collective
excitations by studying the renormalized single-particle properties.
Experimentally, the Angle-resolved photoemission spectroscopy
(ARPES) and scanning tunnelling microscopy (STM) have been powerful
tools for providing the electronic structure and probing the
interaction of the quasiparticle with certain boson modes.

The superconductivity in cuprates can be achieved by doping either
holes or electrons into parent antiferromagnetic (AF) Mott
insulators. One of the most important features in hole doped
cuprates revealed by the ARPES experiments is the slope change of
the quasiparticle dispersion (kink) from the momentum distribution
curve (MDC)~\cite{lan}. The kink is observed along both nodal and
antinodal directions at the energies about $40\sim80$ meV. In the
past few years, the origin of the kinks attracted intensive study
both theoretically and experimentally because it speculated some
kind of interaction which might act as the mysterious glue for
Cooper pairs. Two possible bosonic modes, namely, phonon~\cite{rez}
and spin resonance mode revealed by neutron scattering
experiments~\cite{mook}, have been proposed to account for the
dispersion kink. Theoretically it seems that both electron-phonon
interaction~\cite{dev} and the coupling of the spin resonance
mode~\cite{man,nor,jxli} can reproduce the dispersion kink.
Unfortunately, this two modes could have similar energies, thus it
is difficult to distinguish between the two. Up to now no consensus
has yet been reached. On the other hand, STM experiments also
identified the existence of the bosonic mode in the hole-doped
cuprates, but the origin is also under debate~\cite{lee,pas,dasp}.

In the past few years, more and more attention has been turned to
the electron-doped cuprates. It is well known that the
electron-doped materials exhibit different behaviors from that of
hole-doped ones, namely, they usually have lower superconducting
(SC) transition temperature and narrower SC doping range. Therefore,
the spin resonance energy is much less, namely, only about $10$
meV~\cite{wil,zha} revealed by the neutron scattering experiments.
On the other hand, the phonon energies are expected to be similar to
those of holed-doped ones. Thus the energies of these two modes are
quite different so that their contributions should be easily
separated by experiments. Moreover, earlier ARPES experiments in the
electron-doped cuprates did not observe the kink along the nodal
direction, only the antinodal kink with the energy about $50-70$ meV
was observed~\cite{sato,arm,mat}. Very recently, it was reported by
several groups that the kinks exist in several families of
electron-doped cuprates, both along the nodal
direction~\cite{par,sch,liu,tsu} and antinodal
direction~\cite{par,sch}, with the energy being $50-70$ meV. And the
kinks depend weakly on the doping level and exist even in the normal
state. Another renormalized effect revealed by the experiments is
the peak-dip-hump structure in the energy distribution curve
(EDC)~\cite{sch}, namely, the EDC line-shapes display a sharp
quasiparticle peak near the Fermi energy $E_F$ along the antinodal
direction. This peak terminates and is accompanied with a dip at the
energy about $50$ meV. The peak width decreases when approaching the
Fermi energy. A faint hump-like feature is also revealed in the
nodal direction. Because the kink energy is much greater than the
resonance energy and the spin resonance peak in fact does not exist
in the normal state. Thus it was proposed that the phonon should
account for the dispersion kink~\cite{par,sch,tsu,liu}. On the other
hand, a distinct bosonic mode of the energy about $10$ meV has also
been reported by the STM experiment in electron-doped
cuprates~\cite{nie}. It is proposed that the mode is caused by spin
fluctuations rather than phonons.

In this letter, the spectral function and density of states in
electron-doped cuprates observed by
experiments~\cite{par,sch,tsu,liu,nie} can be reproduced by only
considering the coupling between the spin excitations and the
quasiparticle. We assume phenomenologically that the spin
excitations are from spin fluctuations and the retarded Green's
function $G({\bf k},\omega)$ is a function of the bare normal state
quasiparticle dispersion $\varepsilon_{\bf k}$, the SC
order-parameter $\Delta_{\bf k}$, and the self-energies $\Sigma({\bf
k},\omega)$ due to the coupling of spin fluctuations
~\cite{ram,esc,esc1}. The bare normal state quasiparticle dispersion
is expressed by,
\begin{eqnarray}
\varepsilon_{\bf k}=-2 t_1 (\cos k_x+\cos k_y)-4t_2 \cos k_x\cos k_y
\nonumber\\-2 t_3 (\cos 2k_x+\cos 2k_y)\nonumber\\ -4 t_4 (\cos
k_x\cos 2k_y+\cos k_y \cos 2k_x)\nonumber\\-4 t_5 \cos 2k_x\cos
2k_y-t_0,
\end{eqnarray}
with $t_{0-5}=-82$, $120$, $-60$, $34$, $7$ and $20$ meV. This
single-particle dispersion was used by Ref.~\cite{das} to fit the
ARPES experiments in electron-doped cuprates~\cite{mat}.

The SC order parameter is chosen to have $d$-wave symmetry, namely,
\begin{equation}
\Delta_{\bf k}=\Delta_0(\cos k_x-\cos k_y)/2.
\end{equation}

The spectral function of the electrons can be calculated from the
retarded Green's function as $A({\bf{k}},\omega)=-(1/\pi)$Im$
G_{11}({\bf{k}},\omega+i\delta)$. Here the Green's function $G_{ij}$
($i,j=1,2$) is calculated by Dyson's equation in the Nambu
representation ($2\times2$ matrix), namely,
\begin{equation}\label{g}
\widehat{G}({\bf{k}},\omega+i\delta)^{-1}=\widehat{G}_0({\bf{k}},\omega+i\delta)^{-1}-\widehat{\Sigma}({\bf{k}},\omega+i\delta).
\end{equation}
The bare Green function of the electron $\hat{G}_0$ is expressed by,
\begin{equation}
\hat{G}^{-1}_0({\bf k},\omega)=\left(\begin{array}{cc}
i\omega-\varepsilon_{\bf k}&-\Delta_{\bf k}\\
-\Delta_{\bf k}&i\omega+\varepsilon_{\bf k}
\end{array}\right).
\end{equation}
The self-energy due to spin fluctuation is written
as~\cite{jianxin},
\begin{equation}\label{sig}
\widehat{\Sigma}({\bf{k}},i\omega)=\frac{1}{{\beta}N}\sum_q\sum_{i\omega_m}g^{2}\chi({\bf
q},i\omega_m)
\widehat{\sigma}_3\widehat{G}_0({\bf{k}}-{\bf{q}},i\omega-i\omega_m)\widehat{\sigma}_3,
\end{equation}
where $\widehat{\sigma}_3$ is the Pauli matrix. $\chi({\bf
q},i\omega_m)$ is the spin susceptibility in the random phase
approximation (RPA), namely,
\begin{equation}
\chi({\bf q},\omega)=\frac{\chi_0({\bf q},\omega)}{1+U_{\bf
q}\chi_0({\bf q},\omega)}.
\end{equation}
Here $U_{\bf q}=U_0 (\cos q_x +\cos q_y)$ consistent with the $t-J$
type model. $\chi_0 ({\bf q},\omega)$ can be calculated from the
Fermionic bubble,
\begin{equation}
\chi_0({\bf q},\omega)=-\frac{1}{\beta N} \sum_{{\bf
k},i\omega_m}\mathrm{Tr} [\hat{G}_0({\bf k},i\omega_m)\hat{G}_0({\bf
k}+{\bf q},i\omega+i\omega_m)].
\end{equation}

\begin{figure}
\centering
  \includegraphics[width=2.7in]{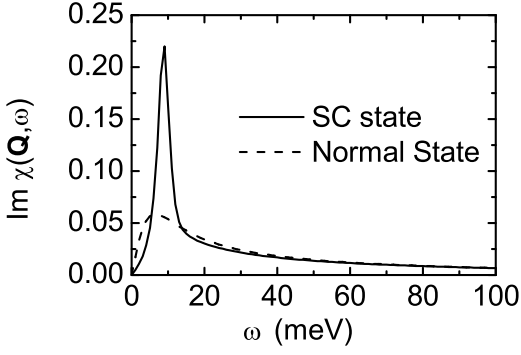}
  \caption{The imaginary part of the spin susceptibility versus the
energy $\omega$ at the AF momentum $Q=(\pi,\pi)$ in the SC state and
normal state, respectively.} \label{fig1}
\end{figure}

In the following presented results, we set $U_0=260$ meV, $g=360$
meV~\cite{note,hkee,aba}. The temperatures and gaps in the SC and
normal states are $T=0.5$ meV, $\Delta_0=10$ meV and $T=T_c=2.3$
meV, $\Delta_0=0$, respectively. We have checked numerically that
the main results are not sensitive to the choice of the parameters.

The imaginary parts of the spin susceptibilities as a function of
the energy $\omega$ are plotted in Fig. 1. A sharp resonance peak is
seen in the SC state at the energy about $\Omega_r=10$ meV. The
origin of the resonance has been studied intensively~\cite{mor,jxl}.
It arises from a collective spin excitation mode corresponding to
the real part of the RPA factor ($1+U_{\bf Q}$Re$\chi_0$) equals to
zero and the imaginary part of the bare spin susceptibility
Im$\chi_0$ is small. The resonance is absent in the normal state,
where the peak intensity decreases dramatically and only a
low-energy broad peak can be seen. While in fact the weight of the
spectra are at low energies and near AF momentum both in the SC and
normal states.

Figs.2(a-d) show the intensity maps of the spectral functions
$[A({\bf k},\omega)f(\omega)]$ ($f(\omega)$ is the Fermi
distribution function) as well as the MDC dispersions in the SC
state (up panels) and normal state (down panels), respectively.
Clear kinks at the energy about $\omega_k\approx 60-70$ meV can be
seen along the antinodal direction. Well defined quasiparticle peaks
exist below the kink energy. At higher energy, the peak intensity is
small. In the normal state the dispersion kink still exists and no
qualitative difference can be seen.
\begin{figure}
\centering
  \includegraphics[width=3.1in]{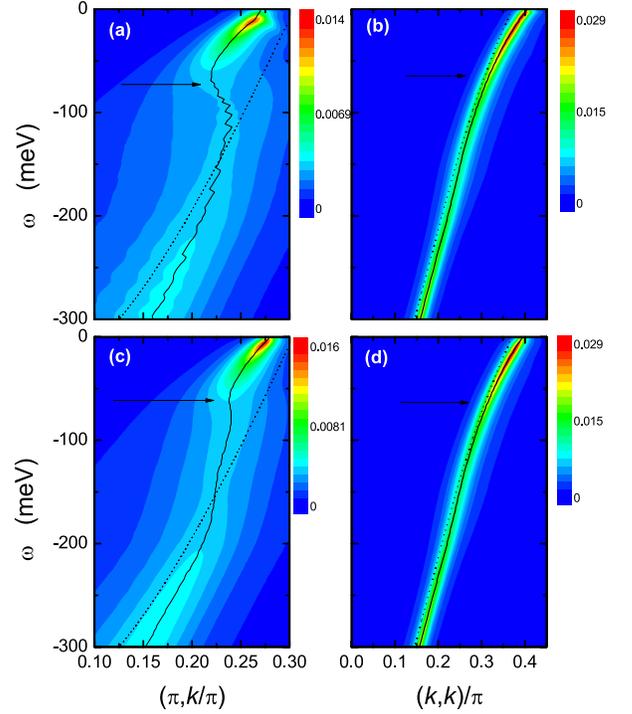}
  \caption{(Color online) The intensity plots of the spectral functions as functions of the momentum and energy
in the SC state (a-b) and normal state (c-d), respectively. The left
and right panels are along $(\pi,0)$ to $(\pi,\pi)$, and $(0,0)$ to
$(\pi,\pi)$ direction, respectively. The solid and dotted lines are
the MDC dispersions and the bare band dispersions, respectively. }
\label{fig2}
\end{figure}

The right panels of Fig.2 show the nodal data of the spectral
function. Here, the dispersion kink can also be seen clearly at the
energy about $\omega_k=60$ meV. The renormalized effect is much
weaker than that of antinodal direction but does exist, which can be
seen more clearly by comparing the dispersion with that of the bare
band. As shown, the renormalized dispersion and the bare one are
nearly parallel at high energies while the renormalized one bends to
low energy at about 60 meV indicating that the kink is indeed caused
by the self-energy. In addition, the peak intensity is larger at low
energies $(\omega \leq \omega_k)$, and decreases evidently at the
kink energy. Similar with the case of antinodal direction, there is
also no remarkable difference between the spectrum of the SC state
with that of the normal state along the nodal direction.

\begin{figure}
\centering
  \includegraphics[width=2.95in]{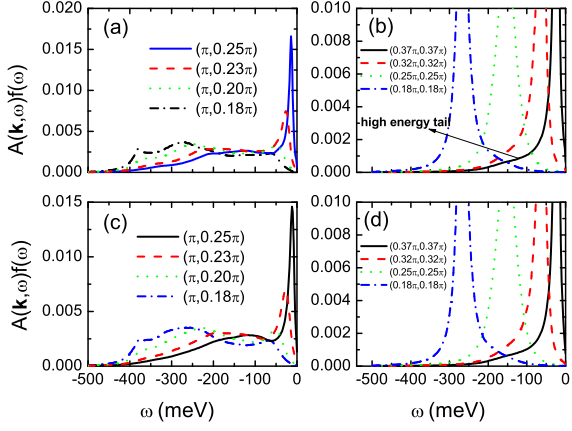}
  \caption{(Color online) The line-shape $A({\bf k},\omega)f(\omega)$ as a function of the energy
at different momentums in the SC state (a-b) and normal state (c-d),
respectively. The left and right panels are along $(\pi,0)$ to
$(\pi,\pi)$, and $(0,0)$ to $(\pi,\pi)$ directions, respectively.}
\label{fig3}
\end{figure}

The EDC line shapes along the antinodal and nodal directions are
plotted in Fig.3. Along the antinodal direction a sharp
quasiparticle peak can be seen near the Fermi momentum $K_F$
following a $50$ meV dip, which is consistent with the experiment as
we mentioned above~\cite{sch}. The peak intensity decreases
dramatically as the momentum is far away from $K_F$. Only a broad
peak can be seen at the momentum $(\pi,0.18\pi)$. Much weaker
renormalized effect is obtained along the nodal direction, namely, a
sharp quasiparticle peak accompanied by a high-energy hump-like tail
near the Fermi energy. At higher energies $(\omega\geq\omega_k)$,
the peak becomes a little broader while it is still well defined and
the hump-like tail disappears. The curve seems to be symmetric with
respect to the peak energy. We can also see that there is no
qualitative difference of the line-shape between the SC state and
normal state.

Our theoretical results reproduce the dispersion kink. Though the
spin susceptibility shows remarkable difference between the SC state
and normal state, namely, a sharp resonance peak can be observed
only in the SC state, as seen in Fig. 1, while the renormalized
spectral function show no evident difference between the SC state
and normal state. Although here we propose that the spin excitations
should be responsible for the kink, while in fact the spin resonance
phenomenon is not essential to the kink. In the following we
demonstrate the origin of the kink and propose that the bare band
structure and the renormalization by the spin susceptibility are
both important to produce the kink.

A sound explanation for the dispersion kink can be given through
analyzing the self-energy. The peak position is determined by the
pole condition $\omega-\varepsilon_{\bf k}-$Re $\Sigma({\bf
k},\omega)=0$ in the normal state. The real-part of the self-energy
is responsible for the kink. Performing the summation over
$i\omega_m$ [Eq.(5)] we can rewrite the self-energy as,
\begin{equation}
\Sigma({\bf k},\omega)=\frac{1}{\pi N}\sum_{\bf
q}{\int}g^{2}\mathrm{Im}\chi({\bf
q},\omega_1)\frac{b(\omega_1)+1-f(\varepsilon_{{\bf k}-{\bf
q}})}{\omega-\omega_1-\varepsilon_{{\bf k}-{\bf
q}}+i\delta}d\omega_1,
\end{equation}
where $b(\omega)$ is the Bose distribution function. The real part
of the self-energy is calculated by using the parameters of the bare
band. The summation over $\sum_{\bf q}$ can be written as the
integral form, $1/(\pi N)\sum_{\bf q}\rightarrow 4\pi \int d {\bf
q}\rightarrow 4\pi\int d\varepsilon_{{\bf k}-{\bf q}} /
[d\varepsilon_{{\bf k}-{\bf q}}/d{\bf q}]$. The spin susceptibility
is peaked at the AF momentum ${\bf Q}$ and very low energy. As a
result, approximately, the absolute of self-energy $\mid \Sigma
\mid$ should have the maxima value and a kink is expected near the
flat band, namely, $\nabla_{\bf k}\varepsilon_{{\bf k}-{\bf Q}}=0$.

We show the real parts of the self-energies in the normal state and
the bare band dispersions along the antinodal direction in Fig.4(a)
and Fig.4(b). As seen in Fig.4(a), the absolute values of the real
part of the self-energy $\mid$ Re $\Sigma$ $\mid$ reach the local
maximum at the energies about $50$ meV and $400$ meV. The origin of
the two peaks can be seen from Fig.4(b), namely, the band dispersion
is flat $(\nabla_{\bf k}\varepsilon_{{\bf k}-{\bf Q}}=0)$ at the
energies about $50$ meV and $400$ meV. As a result, the MDC
dispersion has an obvious kink at the energies about $50-60$ meV
along the antinodal direction. The kink is always there, regardless
Im$\chi({\bf Q},\omega)$ has a resonance peak or not (see Fig. 1).

\begin{figure}
\centering
  \includegraphics[width=3in]{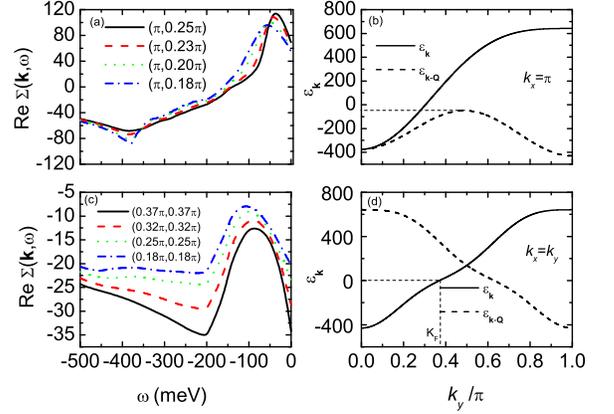}
  \caption{(Color online) The real parts of the self-energies Re $\Sigma({\bf k},\omega)$ vs.
  the energy $\omega$ at different momentums and the bare band
dispersions $\varepsilon_{\bf k}$ and $\varepsilon_{{\bf k}-{\bf
Q}}$ along the antinodal (a-b) and nodal (c-d) directions,
respectively.} \label{fig4}
\end{figure}

The real parts of the self-energies and the bare band dispersions
along nodal directions are shown in Figs.4(c) and (d). As seen from
Fig.4(d), below the Fermi momentum $k<K_F$, $\varepsilon_{{\bf
k}-{\bf Q}}$ is always greater than zero, so that for negative
energies the self-energy is quite small. And different from that
along antinodal direction, there is in fact no obvious peak at low
energies and the absolute value of Re $\Sigma$ is maximum at zero
energy, indicating that the renormalized effect for peak position is
prominent at low energies, consistent with the dispersion shown in
Fig.2. As the momentum is away from the Fermi momentum, the
self-energy tends to be a constant at high energies so that the
renormalized dispersion is parallel to the bare one at high
energies, which can be seen in Fig.2.

For the case of the SC state, because the SC gap is much less than
the kink energy, and the spin susceptibility is still peaked at very
low energy. Thus in fact the SC gap does not influence the kink very
much. We have also check numerously for different gap symmetry,
i.e., the nonmonotonic $d$-wave gap~\cite{mat} and obtain similar
results.

We now turn to address the renormalized effect of the density of
states [$\rho(\omega)=\int A({\bf k},\omega)d{\bf k}$] in the SC
state. Fig.5 shows the density of states as a function of the
energy. The SC coherence peaks at the energies $\pm\Delta_0$ can be
seen clearly. Outside the gaps the hump-like features exist, and
this reveals the existence of the bosonic mode, which is sensitive
to the intensity and energy of the spin resonance peak and located
at the energies about $\pm (\Delta_0+\Omega_r)$ with $\Omega_r\sim
10$ meV, thus the spin resonance mode should account for the humps
outside the gap. This result is consistent with recent STM
experiments on electron-doped cuprates~\cite{nie}. In hole-doped
cuprates similar renormalized effect caused by the bosonic mode was
also predicted theoretically~\cite{zhu} and observed by STM
experiments very recently~\cite{lee,pas,dasp}. We can also see that
the renormalized effect in the electron part (negative energy) and
hole part (positive energy) is asymmetric. The intensity of the SC
coherent peak is smaller and the renormalized hump caused by the
spin resonance is also weaker at the negative energy part. In fact,
the ARPES experiments can only examine the electronic structure in
the electron part so that the possible renormalized effect at the
energy $\Delta_0+\Omega_r$ for spectral function $A({\bf k},\omega)$
is hard to detect and also is not obtained by our calculation. In
fact, very recently the kink at about the energy $20$ meV along
nodal direction was reported by Ref.~\cite{liu} while this result
was not reported by other groups~\cite{par,sch,tsu}.

\begin{figure}
\centering
  \includegraphics[width=2.7in]{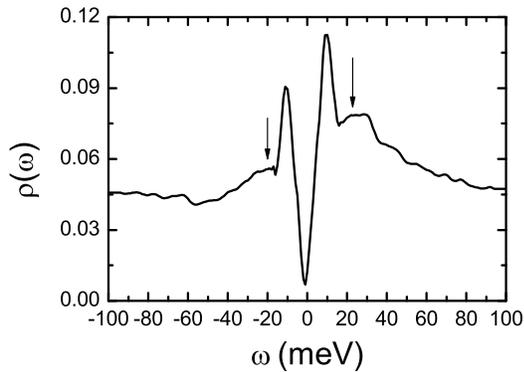}
  \caption{The density of state as a function of the energy in the SC state.} \label{fig5}
\end{figure}

In summary, we study theoretically the effect of the spin
fluctuation mode on the spectral function and density of states in
electron-doped cuprates. We have elaborated that the spin excitation
is able to cause the $50-70$ meV dispersion anomaly and the
hump-like feature of density of states observed by recent
experiments. Thus we present a consistent picture of the effect of
bosonic mode coupling for ARPES and STM spectra in electron-doped
cuprates.

We are grateful to Jian-Xin Zhu and Yan Chen for useful discussions.
This work was supported by the Texas Center for Superconductivity at
the University of Houston and by the Robert A. Welch Foundation
under the Grant no. E-1411.


\begin{thebibliography}{99}
\bibitem{leea} For a review, see, e.g., P. A. Lee, N. Nagaosa, and X. G. Wen, Rev. Mod. Phys. {\bf 78}, 17
(2006).
\bibitem{lan} A. Lanzara $et$ $al$., Nature {\bf 412}, 510
(2001); T. Sato $et$ $al$., Phys. Rev. Lett. {\bf 91}, 157003
(2003); T. K. Kim $et$ $al$., Phys. Rev. Lett. {\bf 91}, 167002
(2003); A. D. Gromko $et$ $al$., Phys. Rev. B {\bf 68}, 174520
(2003); A. A. Kordyuk $et$ $al$., Phys. Rev. Lett. {\bf 92}, 257006
(2004); G.-H. Gweon $et$ $al.$, Nature {\bf 430}, 187 (2004); K.
Terashima $et$ $al$., Nat. Phys. {\bf 2}, 27
(2006); V. B.
Zabolotnyy $et$ $al$., Phys. Rev. Lett. {\bf 96}, 037003 (2006).
\bibitem{rez} D. Reznik $et$ $al.$, Phys.
Rev. Lett. {\bf 75}, 2396 (1995); R. J. McQueeney $et$ $al$., Phys.
Rev. Lett. {\bf 82}, 628 (1999).
\bibitem{mook} H. A. Mook $et$ $al.$, Phys. Rev. Lett {\bf 70}, 3490 (1993).

\bibitem{dev} T. P. Devereaux, T. Cuk, Z.-X. Shen, and N. Nagaosa,
Phys. Rev. Lett. {\bf 93}, 117004 (2004).
\bibitem{man} D. Manske, I. Eremin, and K. H. Bennemann, Phys. Rev. Lett. {\bf
87}, 177005 (2001).
\bibitem{nor} M. Eschrig and M. R. Norman, Phys. Rev. Lett. {\bf
89}, 277005 (2002).
\bibitem{jxli} J. X. Li, T. Zhou, and Z. D. Wang, Phys. Rev. B {\bf
72}, 094515 (2005).
\bibitem{lee} J. Lee $et$ $al.$, Nature {\bf 442}, 546
(2006).
\bibitem{pas} A. N. Pasupathy $et$ $al$., Science {\bf 320},197
(2008).
\bibitem{dasp} P. Das $et$ $al$., Phys. Rev B {\bf 78}, 214505 (2008).
\bibitem{wil} S. D. Wilson $et$ $al$., Nature (London) {\bf 442}, 59
(2006).
\bibitem{zha} J. Zhao $et$ $al$., Phys. Rev. Lett. {\bf 99}, 017001
(2007).

\bibitem{arm} N. P. Armitage $et$ $al$., Phys. Rev B {\bf 68}, 064517
(2003).
\bibitem{mat} H. Matsui $et$ $al$., Phys. Rev. Lett. {\bf 95},
017003 (2005).
\bibitem{sato} T. Sato $et$ $al$., Science {\bf 291}, 1517 (2001).
\bibitem{par} S. R. Park $et$ $al$., Phys. Rev. Lett. {\bf 101},
117006 (2008).
\bibitem{sch} F. Schmitt $et$ $al$., Phys. Rev. B {\bf 78},
100505(R) (2008).

\bibitem{tsu} M. Tsunekawa $et$ $al$., New J. Phys. {\bf 10}, 037005
(2008).
\bibitem{liu} H. Liu $et$ $al$., arxiv: 0808.0802 (unpublished).
\bibitem{nie} F. C. Niestemski $et$ $al$., Nature {\bf
450}, 1058 (2007).
\bibitem{ram} J. Rammer and H. Smith, Rev. Mod. Phys. {\bf 58}, 323 (1986).

\bibitem{esc} M. Eschrig and M. R. Norman, Phys. Rev. Lett. {\bf
85}, 3261 (2000).

\bibitem{esc1} M. Eschrig and M. R. Norman, Phys. Rev. B {\bf 67},
144503 (2003).



\bibitem{das} T. Das, R. S. Markiewicz, and A. Bansil, Phys. Rev. B {\bf 74},
020506(R) (2006).
\bibitem{jianxin} J. X. Li, C. Y. Mou, and T. K. Lee, Phys. Rev.
B {\bf 62}, 640 (2000).
\bibitem{note} The value of the coupling strength $g$ is
controversial, as discussed in Ref.~\cite{hkee,aba}. It is estimated
to be only about $14$ meV in ref.~\cite{hkee} while to be the order
of 1 eV in Ref.~\cite{aba}. The coupling strength we used here is
reasonable according to the $t-J$-type model and Ref.~\cite{aba},
and we have checked numerically that the kink will appear as $g\geq
200$ meV and the kink energy depends weakly on the value of $g$.


\bibitem{hkee} H.-Y. Kee, S. A. Kivelson, and G. Aeppli,
Phys. Rev. Lett {\bf 88}, 257002 (2002).
\bibitem{aba} Ar. Abanov
$et$ $al.$, Phys. Rev. Lett. {\bf 89}, 177002 (2002).

\bibitem{mor} J.-P. Ismer, I. Eremin, E. Rossi, and D. K.
Morr, Phys. Rev. Lett. {\bf 99}, 047005 (2007).
\bibitem{jxl} J. X. Li, J. Zhang, and J. Luo, Phys. Rev. B {\bf 68}, 224503
(2003).
\bibitem{zhu} J. X. Zhu $et$ $al.$, Phys. Rev. Lett. {\bf 92},
017002 (2004).


\end{thebibliography}
\end{document}